\documentclass[prb,twocolumn,superscriptaddress]{revtex4}
\usepackage{multirow}
\usepackage{graphicx}
\usepackage{amsmath}
\usepackage{enumerate}
\usepackage{epstopdf}
\usepackage{times}
\usepackage{color}
\usepackage{bm}
\usepackage{comment}

\definecolor{pink}{rgb}{0.855,0,0.627}

\newcommand{\eto}{Er$_2$Ti$_2$O$_7$}

\begin{document}
\title{Ground State Phase Diagram of Generic ${\bm {XY}}$ Pyrochlore Magnets with Quantum Fluctuations}

\author{Anson W. C. Wong}
\affiliation{Department of Physics and Astronomy, University of British Columbia, Vancouver, BC, V6T 1Z1, Canada}
\affiliation{Department of Physics and Astronomy, University of Waterloo, Waterloo, ON, N2L 3G1, Canada}
\author{Zhihao Hao}
\affiliation{Department of Physics and Astronomy, University of Waterloo, Waterloo, ON, N2L 3G1, Canada}
\author{Michel J. P. Gingras}
\affiliation{Department of Physics and Astronomy, University of Waterloo, Waterloo, ON, N2L 3G1, Canada}
\affiliation{Perimeter Institute for Theoretical Physics, 31 Caroline North, Waterloo, ON, N2L 2Y5, Canada}
\affiliation{Canadian Institute for Advanced Research, 180 Dundas Street West, Suite 1400, Toronto, ON, M5G 1Z8, Canada}

\date{\today}

\begin{abstract}
Motivated by recent experimental and theoretical progress on the Er$_2$Ti$_2$O$_7$ pyrochlore $XY$ 
antiferromagnet, we study the  problem of quantum order-by-disorder in pyrochlore $XY$ systems.
We consider the most general  nearest-neighbor pseudo spin-1/2 Hamiltonian for such a system characterized by anisotropic
spin-spin couplings $J_e\equiv$$\{ J_\pm, J_{\pm\pm}, J_{z\pm}, J_{zz}\} $ and construct zero-temperature phase diagrams.
Combining  symmetry arguments and spin-wave calculations, we show that the ground state phase boundaries
 between the two candidate ground states of the $\Gamma_5$ irreducible representation,
the $\psi_2$ and $\psi_3$ (basis) states,  are rather accurately determined by a cubic equation in 
$(J_{\pm}J_{\pm\pm})/J_{z\pm}^2$.
Depending on the value of $J_{zz}$, there can be one or three phase boundaries that separate alternating 
regions of $\psi_2$ and $\psi_3$ states.
In particular, we find for sufficiently small $J_{zz}/J_{\pm}$ a narrow $\psi_2$ sliver sandwiched between 
two $\psi_3$ regions in the
$J_{\pm\pm}/J_\pm$ vs $J_{z\pm}/J_\pm$ phase diagram.
From our results, one would be able to predict which state ($\psi_2$ or $\psi_3$) may be realized in
a real material given its set of $J_e$ couplings.
Our results further illustrate the very large potential sensitivity of the ground state of $XY$ pyrochlore systems  to minute  changes in their spin Hamiltonian.
Finally, using the experimentally determined 
$J_e\equiv$$\{ J_\pm, J_{\pm\pm}, J_{z\pm}, J_{zz}\} $ and $g$-tensor values for Er$_2$Ti$_2$O$_7$, 
we show that the heretofore neglected long-range $1/r^3$ magnetostatic dipole-dipole interactions do not change the conclusion that 
Er$_2$Ti$_2$O$_7$ has a $\psi_2$ ground state induced via a quantum order-by-disorder mechanism.
As a new avenue of research in $XY$ pyrochlore materials distinct from the rare-earth pyrochlore oxides,
we propose that the Cd$_2$Dy$_2$Se$_4$ chalcogenide spinel,  
in which the Dy$^{3+}$ ions form a pyrochlore lattice and may be $XY$-like,  could prove interesting to investigate.

\end{abstract}
\maketitle

\section{Introduction}
Simplified Hamiltonian (${\cal H})$ models of magnetic systems with competing or geometrically frustrated interactions often feature
 a large number of accidentally degenerate classical ground states.
Such a degeneracy can typically be lifted energetically by additional perturbations to ${\cal H}$, such as further neighbor interactions,
 magnetic anisotropy as well as spin-lattice coupling \cite{LMM}.
A more exotic mechanism is one in which thermal or quantum fluctuations induce long-range order within the degenerate manifold. 
This is the thermal or quantum order-by-disorder (ObD) mechanism \cite{Villain.1980,Shender.1982,Henley.PhysRevLett.62.2056,Yildirim.1999}. 

While thermal \cite{Villain.1980} and  quantum \cite{Shender.1982} ObD has been proposed to be at play in a number of condensed matter systems,
the number of compelling experimental demonstrations of ObD among real materials, as opposed to theoretical models, has remained quite limited.
In this context, the Er$_2$Ti$_2$O$_7$ insulating magnetic pyrochlore oxide \cite{Gardner_RMP} 
stands
as a promising textbook example where
ObD is at the origin of the experimentally observed long-range order.

\begin{figure}
\centering
\includegraphics[width=0.45\columnwidth]{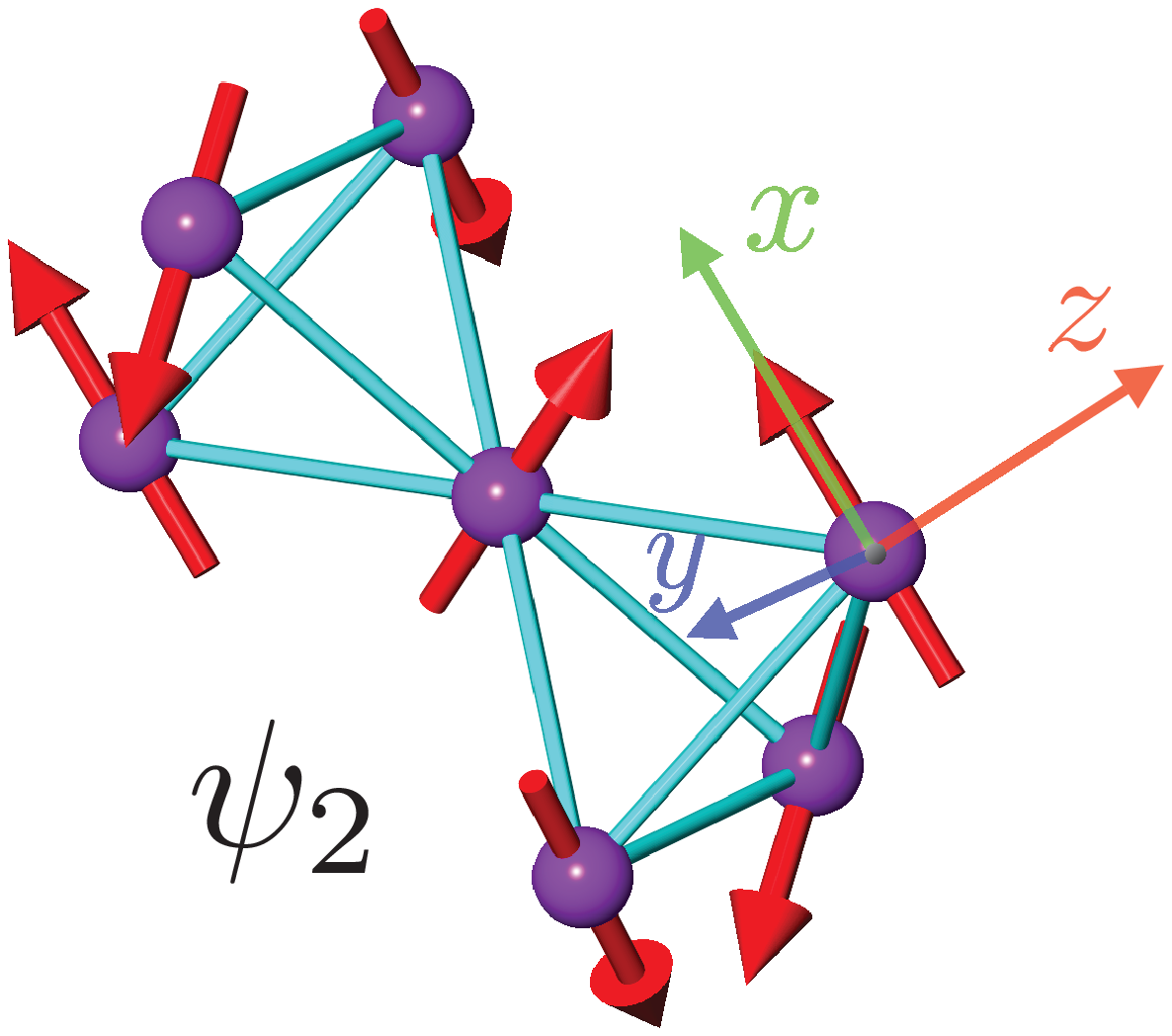}\qquad\includegraphics[width=0.45\columnwidth]{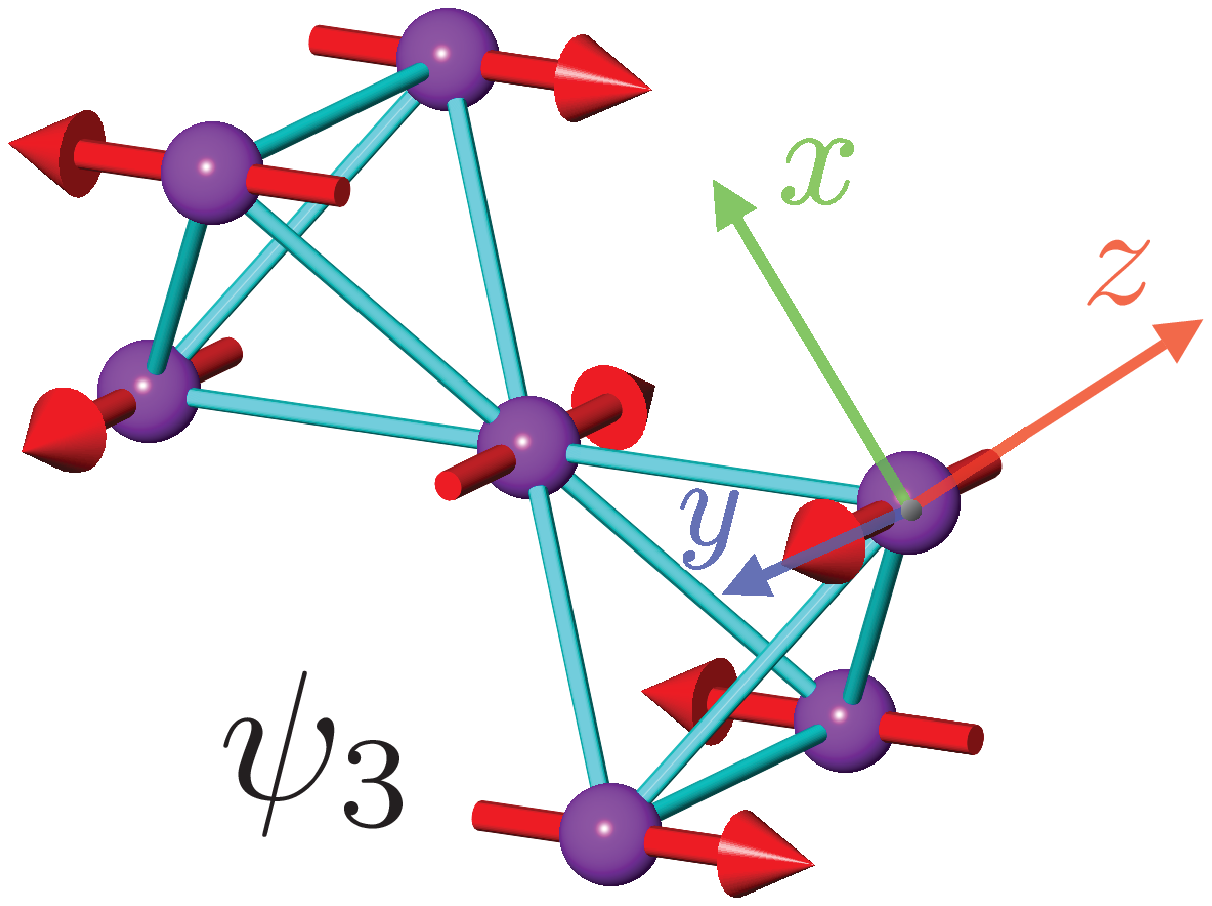}
\caption{Spin configurations of $\psi_2$ (left) and $\psi_3$ (right) states.
See main text for description of the spin orientation for each state.}
\label{psi2and3}
\end{figure}

In Er$_2$Ti$_2$O$_7$,  the magnetic Er$^{3+}$ ions form a three-dimensional network of corner-sharing tetrahedra, the so-called ``pyrochlore'' lattice \cite{Gardner_RMP}. 
A free Er$^{3+}$ ion has angular momentum $J=15/2$. The $^4$I$_{15/2}$ multiplet splits in the local crystal-field environment, 
yielding a Kramers doublet as  lowest energy levels \cite{Champion.PhysRevB.68.020401}.  The resulting pseudo spin-$1/2$ moment describing this magnetic doublet 
lies preferentially in an $xy$ plane (see Fig. \ref{psi2and3}) 
perpendicular to the three-fold symmetry $\hat z$ axis at the ion site,  which is one of the cubic $[111]$ directions. 
Er$_2$Ti$_2$O$_7$ undergoes a second order transition to long-range order at a critical
 temperature $T_c\approx 1.2$ K \cite{Blote,Champion.PhysRevB.68.020401,Ruff_ETO,Sosin_ETO,Dalmas_ETO}.
Defining $\phi_{\bm r}$ as the azimuthal angle of the magnetic moment
in the $xy$ plane expressed in specific  sublattice coordinates \cite{Ross.2011} for site ${\bm r}$,  
neutron scattering studies \cite{Champion.PhysRevB.68.020401,Poole.2007} determined that Er$_2$Ti$_2$O$_7$ orders in a state with 
$\phi_{\bm r}=\phi=n\pi/3$ ($n=0,1,2\ldots$), the so-called $\psi_2$ state (see Fig. \ref{psi2and3}, left),
 in contrast to the $\psi_3$ state with $\phi=\pi/2+n\pi/3$ (see Fig. \ref{psi2and3}, right). 
Both $\psi_2$ and $\psi_3$  are basis states of the 
$\Gamma_5$ irreducible representation (irrep.) \cite{Poole.2007}, referred to as the
$\Gamma_5$ manifold hereafter.
While it had  been recognized for some time that quantum ObD might be at 
work in Er$_2$Ti$_2$O$_7$ \cite{Champion.PhysRevB.68.020401,Champion.2004}, 
the details via  which ObD does actually operate and select the non-coplanar $\psi_2$ state had remained 
a central unsolved problem until recently
\cite{McClarty_ETO,Stasiak.2011,Zhitomirsky.PhysRevLett.109.077204,Savary.PhysRevLett.109.167201}.

Much progress has been achieved over the past year \cite{Zhitomirsky.PhysRevLett.109.077204,Savary.PhysRevLett.109.167201} 
 in reaching a resolution of this quandary.
Building on previous works \cite{McClarty_ETO,Stasiak.2011}, 
Refs.~[\onlinecite{Zhitomirsky.PhysRevLett.109.077204,Savary.PhysRevLett.109.167201}]
showed that quantum ObD can select a $\psi_2$ state in effective spin-1/2 models
with anisotropic exchange interactions and relevant to Er$_2$Ti$_2$O$_7$.
In particular, Savary \emph{et al.} \cite{Savary.PhysRevLett.109.167201} obtained a microscopic pseudo spin-1/2 
model of Er$_2$Ti$_2$O$_7$ by fitting the spin-wave spectrum of a field-polarized paramagnetic 
state measured using inelastic neutron scattering. 
These authors pointed  out that no mechanism exists at the mean-field level
 via which arbitrary range anisotropic interactions can lift the degeneracy between 
$\psi_2 $ and $\psi_3$ \cite{but_excited_CEF}. 
On the basis of their Er$_2$Ti$_2$O$_7$-parametrized model, Ref.~[\onlinecite{Savary.PhysRevLett.109.167201}]
showed that quantum fluctuations select a $\psi_2$ ground state as
 observed  in Er$_2$Ti$_2$O$_7$  \cite{Champion.PhysRevB.68.020401,Poole.2007}. 
We show below that including long-range  dipolar interactions does not
change this conclusion.

These recent  developments have motivated us to study the following  question: 
\emph{Given a pyrochlore $XY$ magnet where the 
$\Gamma_5$ manifold has the lowest classical energy, which state, $\psi_2$ or $\psi_3$, 
is selected by quantum fluctuations at temperatures close to zero}? 
In addition to its broad relevance to the field of frustrated magnetism,
this question is of importance to help understand  the ground state
displayed by real magnetic pyrochlore compounds \cite{Gardner_RMP}.

To address this question, we investigate the most general symmetry-allowed Hamiltonian of Eq.~\eqref{hami} below
\cite{Ross.2011,Savary.PhysRevLett.108.037202} with nearest-neighbor bilinear spin interactions on the pyrochlore lattice.
For the interaction parameters where the ground state belongs to the $\Gamma_5$ manifold, 
we determine the phase boundaries of the $\psi_2$ and $\psi_3$ states by comparing the contribution of 
the quantum zero-point energy from spin-waves.  Combining the spin-wave results with  symmetry arguments, 
we show that, depending onthe value of the anisotropic couplings,  {\it either} 
$\psi_2$ or $\psi_3$  can be selected by quantum fluctuations.  
This is consistent with previous Monte-Carlo studies \cite{Stasiak.2011,Chern.2010,Zhitomirsky.PhysRevLett.109.077204}
for specific sets of interaction parameters where the ground state was 
found to be either $\psi_2$ or $\psi_3$.

The rest of the paper is organized as follows. In Sec. \ref{mands}, we present the model
 and describe the main features of its phase diagram based on symmetry arguments. 
We use spin wave calculations to map out the phase diagram quantitatively in Sec. \ref{sec:sw}.  
We find that Er$_2$Ti$_2$O$_7$  sits deep in the $\psi_2$ part of the phase diagram.
 In Sec. \ref{sec:dp}, we show that long-range dipolar interaction does 
not change the $\psi_2$ ground state selection in \eto. 
Finally, Sec. \ref{sec:discuss} concludes our paper
and discusses future directions, proposing in particular that  CdDy$_2$Se$_4$ 
could be an interesting material to explore the physics of $XY$ pyrochlore magnets outside the realm of the 
$R_2M_2$O$_7$ pyrochlore oxides \cite{Gardner_RMP}.

\section{Model and symmetry considerations}\label{mands}
We consider the  effective $S=1/2$ spin Hamiltonian
${\cal H} = {\cal H}_0 + {\cal H}_1$
on the pyrochlore lattice with anisotropic nearest-neighbor exchange couplings 
$J_e \equiv \{ J_\pm, J_{\pm\pm}, J_{z\pm}, J_{zz} \}$:\cite{Ross.2011}:
${\cal H}={\cal H}_0+{\cal H}_1$ \cite{Ross.2011}:
\begin{subequations}
\label{hami}
\begin{eqnarray}
{\cal H}_0 &=&-J_{\pm}\sum_{\langle ij\rangle}(S_{i}^{+}S_{j}^{-}+h.c.)+J_{zz}\sum_{\langle ij\rangle}S_{i}^{z}S_{j}^{z}\\
{\cal H}_1 &=&\sum_{\langle ij\rangle}J_{\pm\pm}(\gamma_{ij}S_{i}^{+}S_{j}^{+}+\gamma^\ast_{ij}S_{i}^-S_j^{-})+\nonumber\\
&&J_{z\pm}\left(S_{i}^{z}(\zeta_{ij}S_j^{+}+\zeta^\ast_{ij}S_j^{-})+i\leftrightarrow j\right) ,
\end{eqnarray}
\end{subequations}
where $\gamma_{ij}=-\zeta_{ij}^*$ are bond-dependent phases \cite{Ross.2011,Savary.PhysRevLett.108.037202} (Appendix \ref{AA}).  
All spin components are written in terms of local coordinates (see Fig. \ref {psi2and3}).
If ${\cal H}_1$ is absent and $J_{zz}=0$,  ${\cal H}$ in Eq. \eqref{hami} reduces to a ferromagnetic $XY$ model 
(in terms of spin components in the local coordinates) for which the $XY$ ($U(1)$) symmetry is exact. 
In other words, ${\cal H}_0$ is invariant under a simultaneous rotation about all local $z$ ($[111]$) axes.
As a result, the $\psi_2$ and $\psi_3$ states have the same classical ground state energy.
We observe that $J_{\pm}$ is the primary interaction that favors $\Gamma_5$  as lowest energy manifold
and note, incidentally, that $J_\pm$ is the largest coupling in Er$_2$Ti$_2$O$_7$ \cite{Savary.PhysRevLett.109.167201}.
In the remainder of the paper we set $J_{\pm}=1$, with all energies henceforth measured in units of $J_{\pm}$,
denoting the scaled interactions by the corresponding lower-case letters, $j_{zz}\equiv J_{zz}/J_{\pm}$ for example. 

A finite $j_{zz}$ does not break the $U(1)$ symmetry and the consequential Ising couplings 
$J_{zz}S_i^{z}S_j^{z}$ are thus part of ${\cal H}_0$ in Eq. \eqref{hami}.
The character of the ground state changes, however, from $XY$-like to Ising-like 
if the magnitude of $j_{zz}$ exceeds certain critical values:  
Two-in/two-out spin-ice states \cite{Bramwell.1998} 
are the lowest energy states for $j_{zz}>6$ while the ground state becomes the all-in-all-out \cite{Bramwell.1998} state if $j_{zz}<-2$. 

The  lattice is not invariant under an arbitrary rotation about the local $z$ ($[111]$ cubic) axes. 
This allows for a finite ${\cal H}_1$ that breaks the $U(1)$ symmetry \emph{explicitly} \cite{Savary.PhysRevLett.109.167201}.
 In addition to Ising-like states, the $\Gamma_5$ manifold is adjacent to several other phases 
in the parameter space spanned by $j_{z\pm}$, $j_{\pm\pm}$ and $j_{zz}$. For $j_{\pm\pm}>2$, 
the Palmer-Chalker (PC) state \cite{palmer.PhysRevB.62.488} (i.e. the $\psi_4$ state of the $\Gamma_7$ irrep.
 \cite{Poole.2007}) is the ground state. Simple energy minimization determines 
the phase boundary between the $\Gamma_5$ 
manifold and a state with ferromagnetic moments canted from the $[100]$ cubic direction, 
or ``splayed ferromagnetic'' (SF) state \cite{Savary.PhysRevLett.108.037202,Yaouanc.PhysRevLett.110.127207}.
\begin{equation}
\label{boundary}
j_{\pm\pm}^{(\Gamma_5)}=\frac{4j_{z\pm}^2}{6-j_{zz}}-2  . 
\end{equation}
The $\Gamma_5$ with degenerate $\psi_2$ and $\psi_3$ states is realized in a range of $j_{\pm\pm}$ that satisfies
$j_{\pm\pm}^{(\Gamma_5)} < j_{\pm\pm} < 2$.

Within $\Gamma_5$,  the energy of $\psi_2$ and $\psi_3$ remains the same at the classical level 
for arbitrary $( J_\pm, J_{\pm\pm}, J_{z\pm},J_{zz} 	)$ \cite{Savary.PhysRevLett.109.167201}.  
This \emph{accidental} degeneracy is expected to be generically lifted by quantum fluctuations.
To lowest order, the energy contribution from quantum fluctuations is the sum of the zero-point energy of spin-wave modes, 
 denoted as $E_{0}(\psi_{i})$ with $i=2,3$. Before describing the results of our spin-wave calculations, 
it is instructive to investigate the expected analytical properties of 
$\delta E(j_{\pm\pm},j_{z\pm})\equiv E_{0}(\psi_2)-E_{0}(\psi_3)$ on the basis of a symmetry analysis.

 Consider first $\delta E(j_{\pm\pm},0)$; a rotation about the local $z$ axes by $\pi/2$ transforms $\phi\to\phi+\pi/2$.
 Consequentially,  $j_{\pm\pm}$ transforms as $j_{\pm\pm}\to j_{\pm\pm}\exp(i\pi)=-j_{\pm\pm}$ 
while the spin configurations of the $\psi_2$ and $\psi_3$ states are interchanged:
 \begin{equation}\label{odd}
 \begin{aligned}
 \delta E(j_{\pm\pm},0)&\to \delta E(-j_{\pm\pm},0)=E_0(\psi_3)-E_0(\psi_2)\\&=-\delta E(j_{\pm\pm},0). 
 \end{aligned}
 \end{equation}
We conclude that $\delta E(j_{\pm\pm},0)$ is an odd function of $j_{\pm\pm}$.
 Similarly,  a rotation about the local $z$ axes by $\pi$ changes $\phi\to\phi+\pi$.
Under this transformation, $j_{z\pm}\to j_{z\pm}\exp(i\pi)=-j_{z\pm}$ and $j_{\pm\pm}\to j_{\pm\pm}$, with
the $\psi_2$ and $\psi_3$ states preserved under this rotation, and 
$\delta E(j_{\pm\pm},j_{z\pm})$ is thus an even function of $j_{z\pm}$. 
These symmetry properties constrain the overall topology of the phase diagram for $-2<j_{zz}<6$.

 \begin{figure}[t!]
 \centering
 \includegraphics[width=0.90\columnwidth]{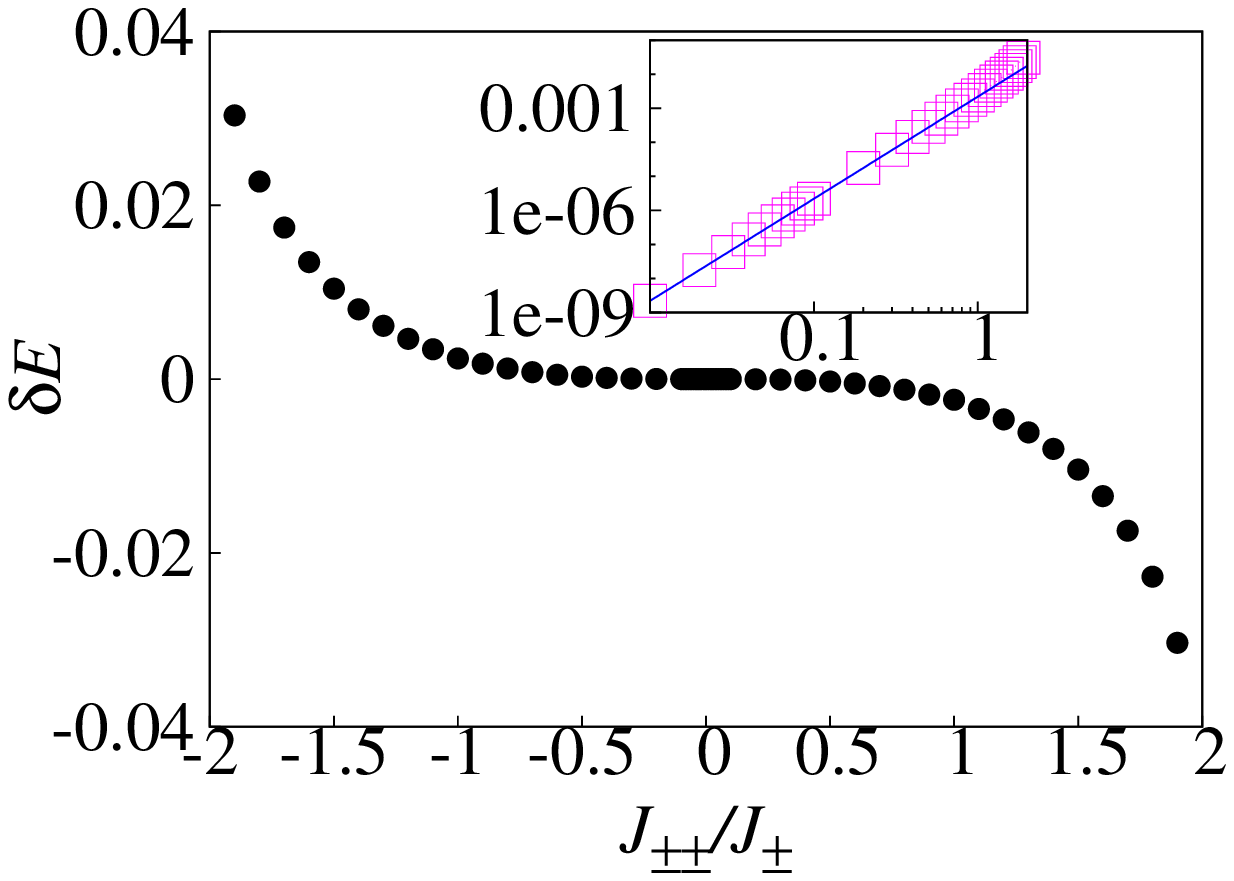}
 \includegraphics[width=0.90\columnwidth]{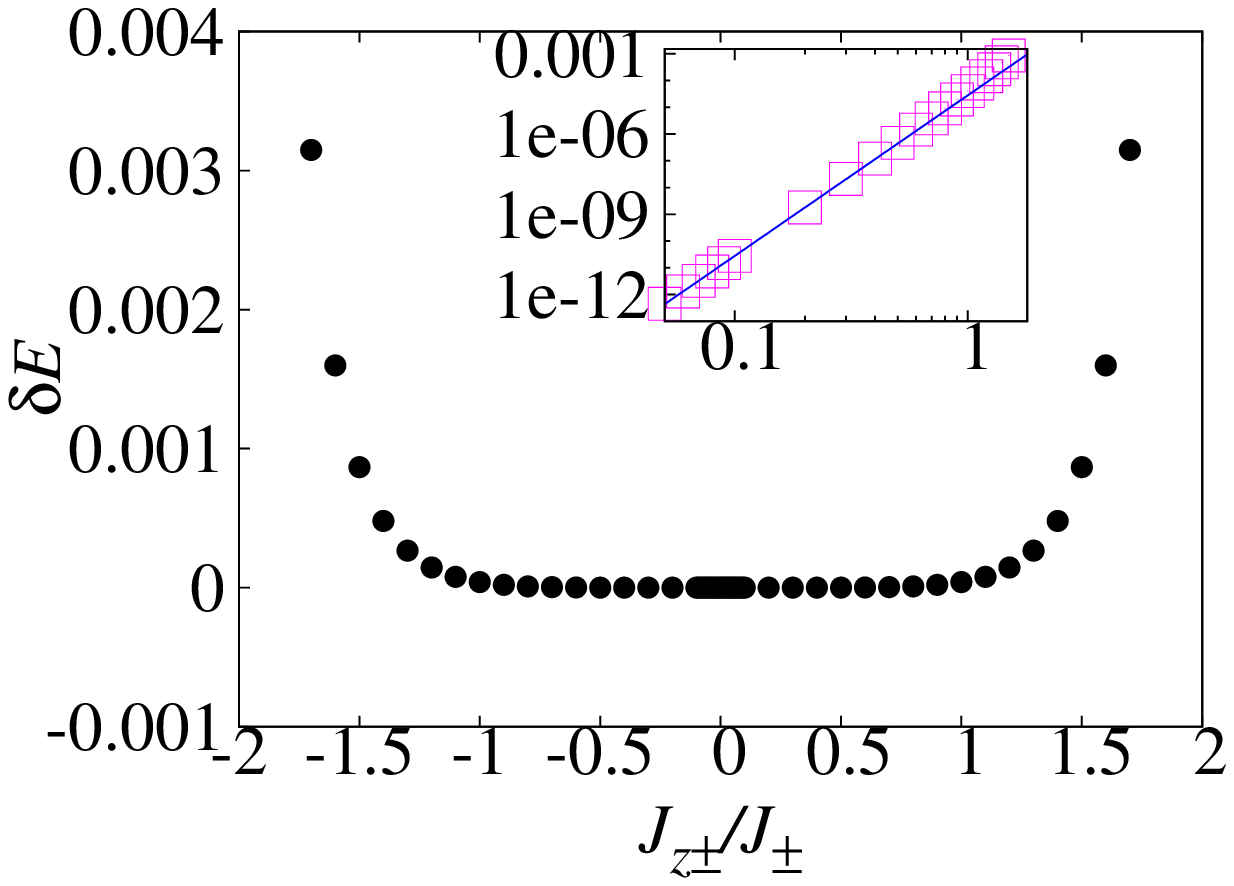}
 \caption{$\delta E(j_{\pm\pm},0)$ (top) is an odd function of $j_{\pm\pm}$ 
and $\delta E(0,j_{z\pm})$ (bottom) is an even function of $j_{z\pm}$.
The insets show, in the form of log-log plots,  the fit (solid blue line)
of $|\delta E(j_{\pm\pm},0)|$ to $j_{\pm\pm}^3$ and $\delta E(0,j_{z\pm})$ 
to $j_{z\pm}^6$ for $j_{\pm\pm}>0$ and $j_{z\pm}>0$,  respectively.
The fits (blue solid lines) were performed over the range $[0.01,0.1]$ for both $j_{\pm\pm}$ and $j_{z\pm}$.
The deviation of the data (symbols) from the fit reflects the slight
departure of $\delta E(\phi)$ from the strict $\cos(6\phi)$
form and thus the cubic polynomial in $x$ in Eq.~\eqref{cubic}  .
}
\label{linescan}
 \end{figure}

Symmetry considerations also help us write down the lowest order expansion of $\delta E$ in terms of 
a polynomial in the dimensionless couplings $j_{z\pm}$ and $j_{\pm\pm}$. Due to the cubic symmetry of the lattice, 
the lowest order and most relevant $U(1)$ symmetry-breaking term is $\cos(6\phi)$ \cite{McClarty_ETO,Savary.PhysRevLett.109.167201} 
where $\phi$ is the azimuthal angle in the $xy$ plane of the local frame. $\cos(6\phi) =+1$ and $-1$ 
for the $\psi_2$ and $\psi_3$ state, respectively, and 
$\delta E$ is proportional 
to the coefficient of $\cos(6\phi)$. We note that anisotropy terms 
of the form $\sin(6n\phi)$ are absent: using the definition of local $z$ axis in Fig. \ref{psi2and3}, 
an improper four-fold rotation around the global $z$ axis together with a time reversal 
transforms $S_{i}^{x}$ to $S_i^x$ and
           $S_i^y$ to $-S_{i}^{y}$, or $\phi$ to $-\phi$.
Since $\sin(6n\phi)$ is odd 
 under this transformation, such terms are forbidden in the (free) energy. 

Since one power of $S^{+}$ contributes $\exp(i\phi)$,
we write down the polynomial expansion of $\delta E$ with the help of a  simple power counting relation, 
$j_{\pm\pm}\sim \exp(\pm i2\phi)$ and $j_{z\pm} \sim \exp(\pm i\phi)$, getting:
\begin{equation}
\label{cubic}
\begin{aligned}
\delta E&\approx c_3 j_{\pm\pm}^3+c_2j_{\pm\pm}^2j_{z\pm}^2+c_1 j_{\pm\pm}j_{z\pm}^4+c_0j_{z\pm}^6\\
&\equiv c_3j_{z\pm}^6\left(x^3+\tilde{c}_2 x^2+\tilde{c}_1 x+\tilde{c}_0\right),
\end{aligned}
\end{equation}
where $x\equiv j_{\pm\pm}/j_{z\pm}^2$. and the $c_i$ coefficients are  functions of $j_{zz}$. 
Restoring all factors of $J_{\pm}$, we have $x=(J_{\pm}J_{\pm\pm})/J_{z\pm}^2$. 
 The form \eqref{cubic} is the most general cubic polynomial in $j_{\pm\pm}$ and $j_{z\pm}^2$ and
the phase boundaries between the $\psi_2$ and $\psi_3$ states are determined by the 
real solutions 
of  Eq.~\eqref{cubic} with $\delta E=0$  .
We therefore expect the number of phase boundaries 
between $\psi_2$ and $\psi_3$ states to evolve smoothly between $1$ and $3$ as $j_{zz}$ is varied.
 We now proceed to explicitly check these expectations on the basis of spin-wave calculations about  the $\psi_2$ and $\psi_3$ states. 

\begin{figure*}[ht]
\begin{centering}

\noindent\begin{minipage}{\textwidth}

%A
\noindent\begin{minipage}[b]{.33\textwidth}
\includegraphics[width=6cm]{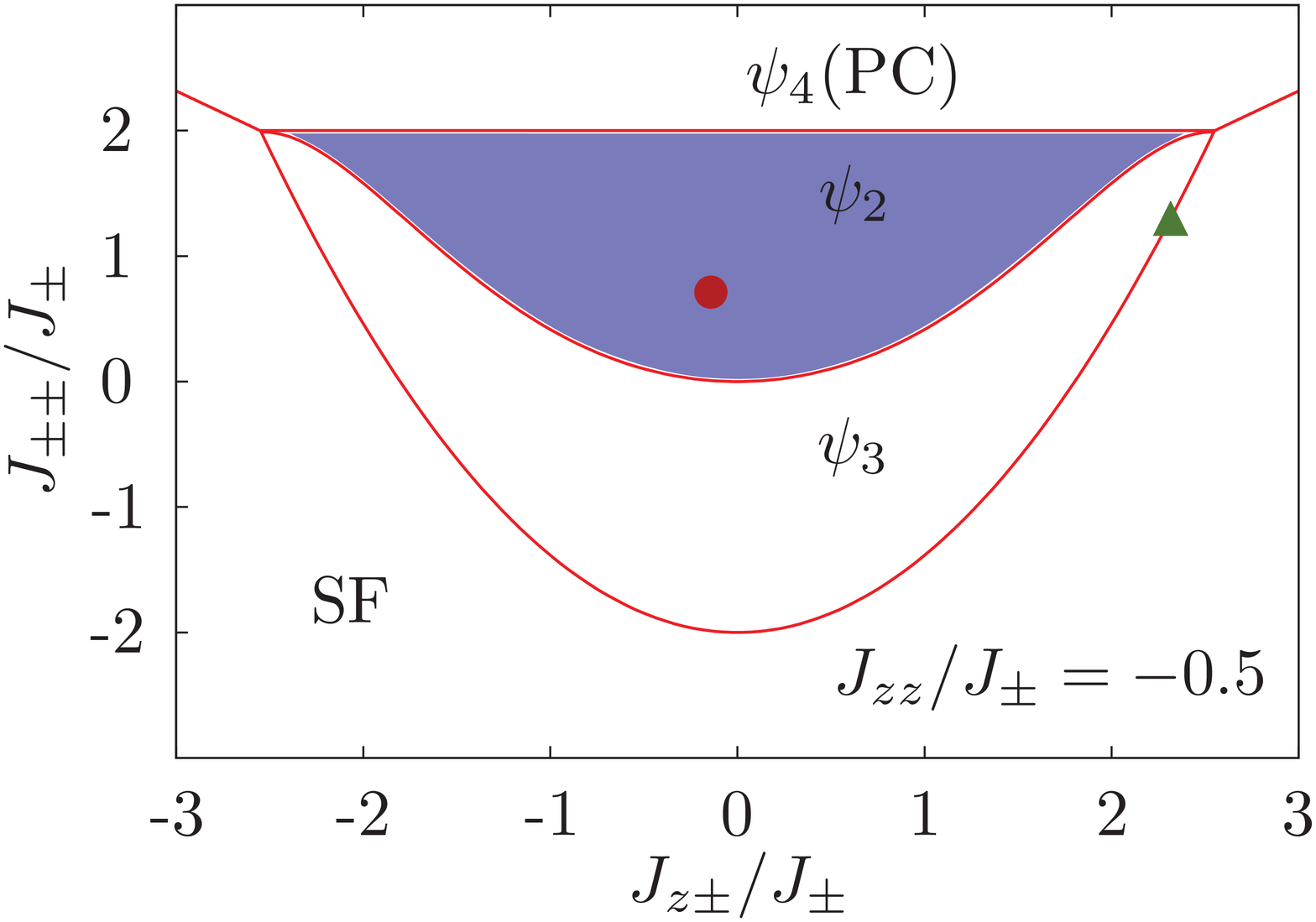}
\end{minipage} 
\hfill
\begin{minipage}[b]{.33\textwidth}
\includegraphics[width=6cm]{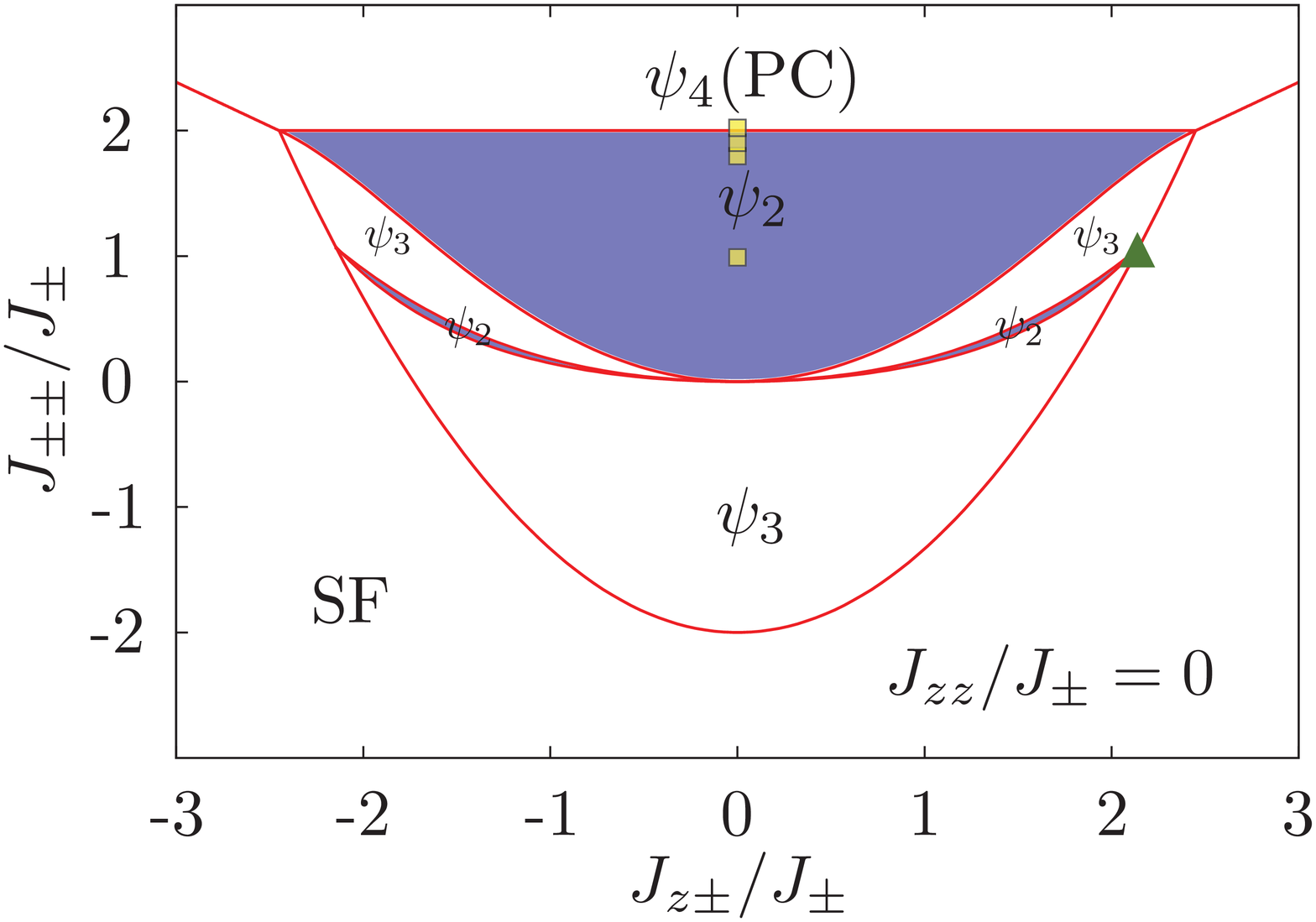}
\end{minipage}
\hfill
\begin{minipage}[b]{.33\textwidth}
\includegraphics[width=6cm]{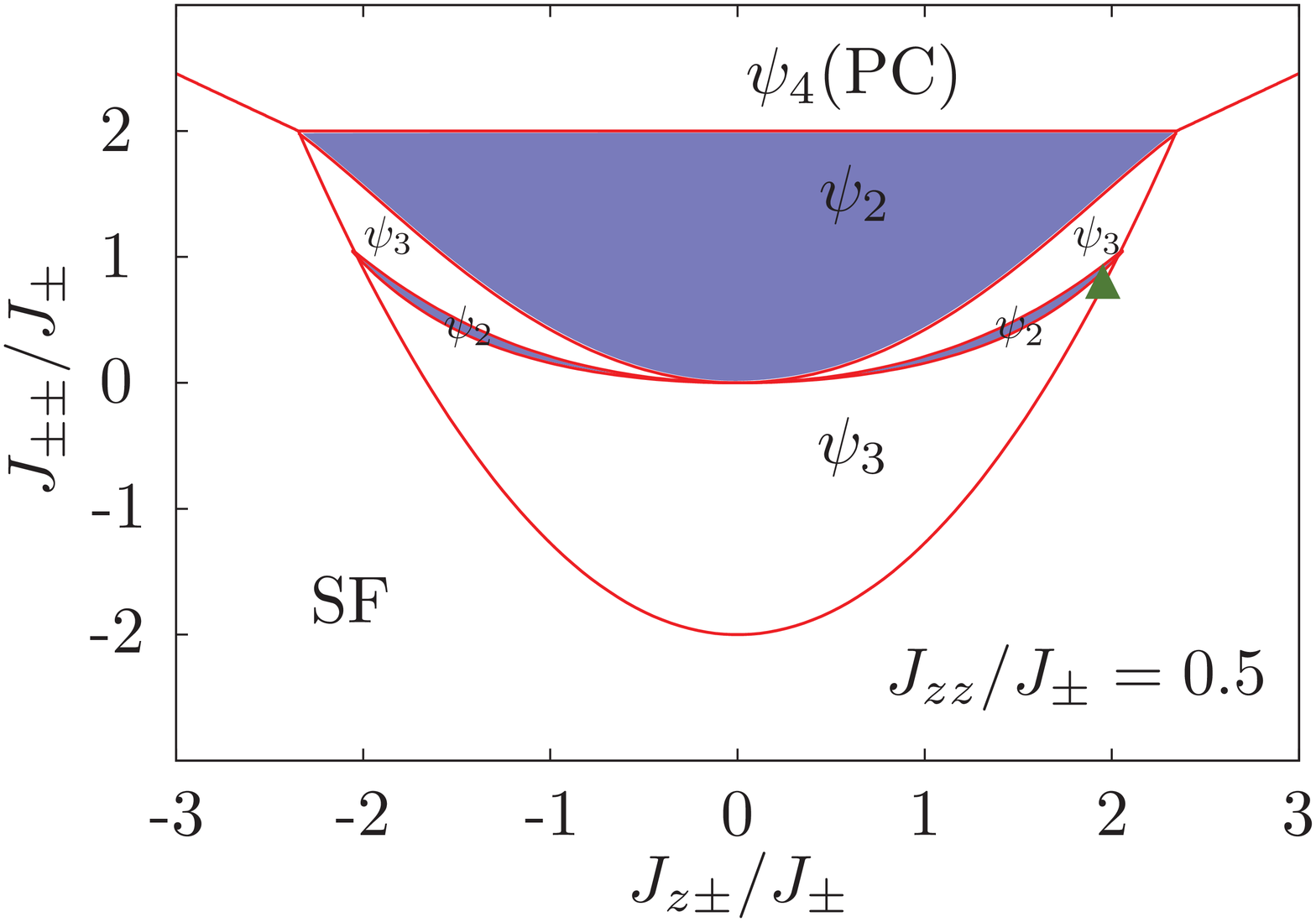}
\end{minipage}

\end{minipage}

\vspace{2ex}

%B
\noindent\begin{minipage}{\textwidth}

\noindent\begin{minipage}[b]{.33\textwidth}
\includegraphics[width=6cm]{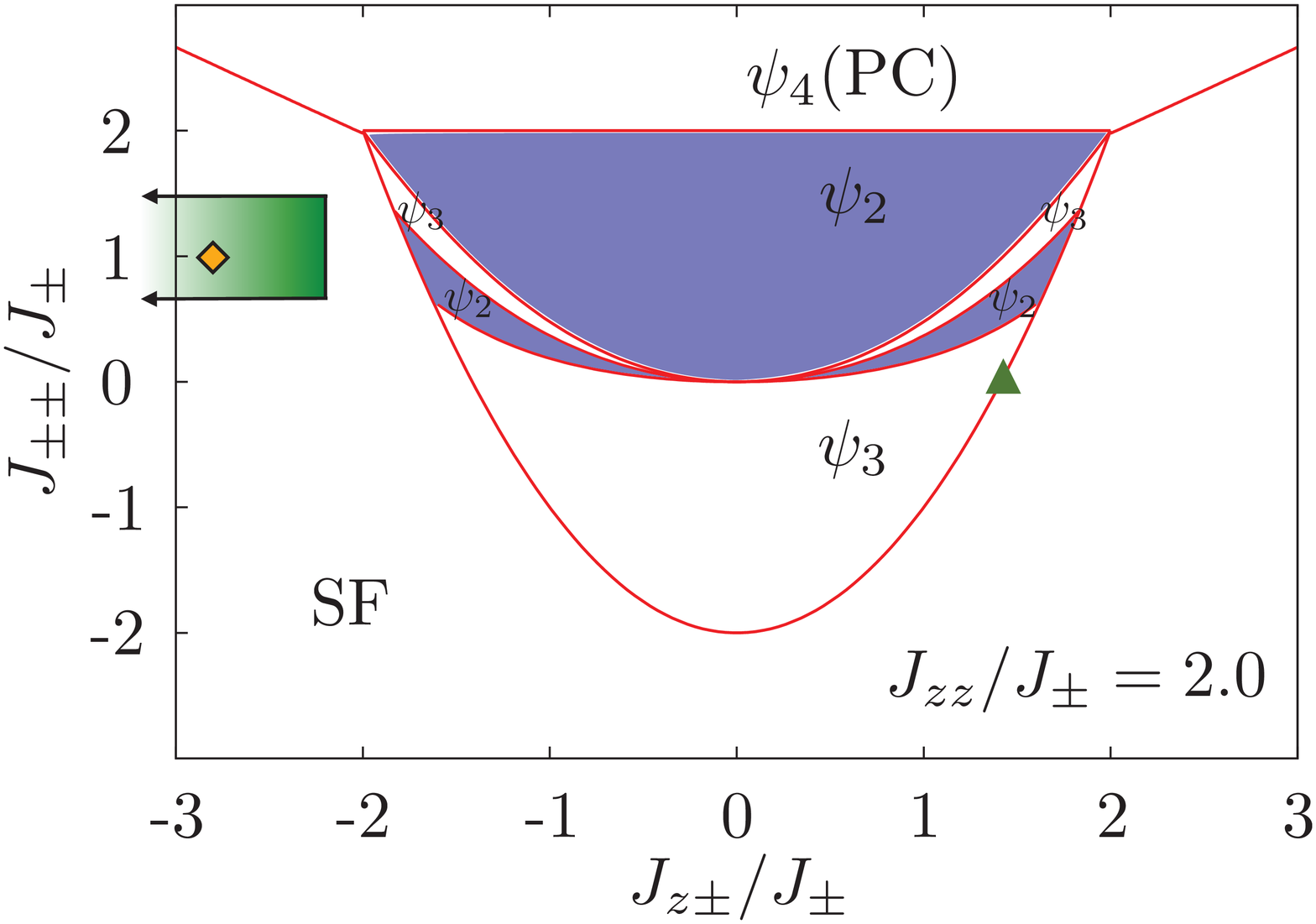}
\end{minipage} 
\hfill
\begin{minipage}[b]{.33\textwidth}
\includegraphics[width=6cm]{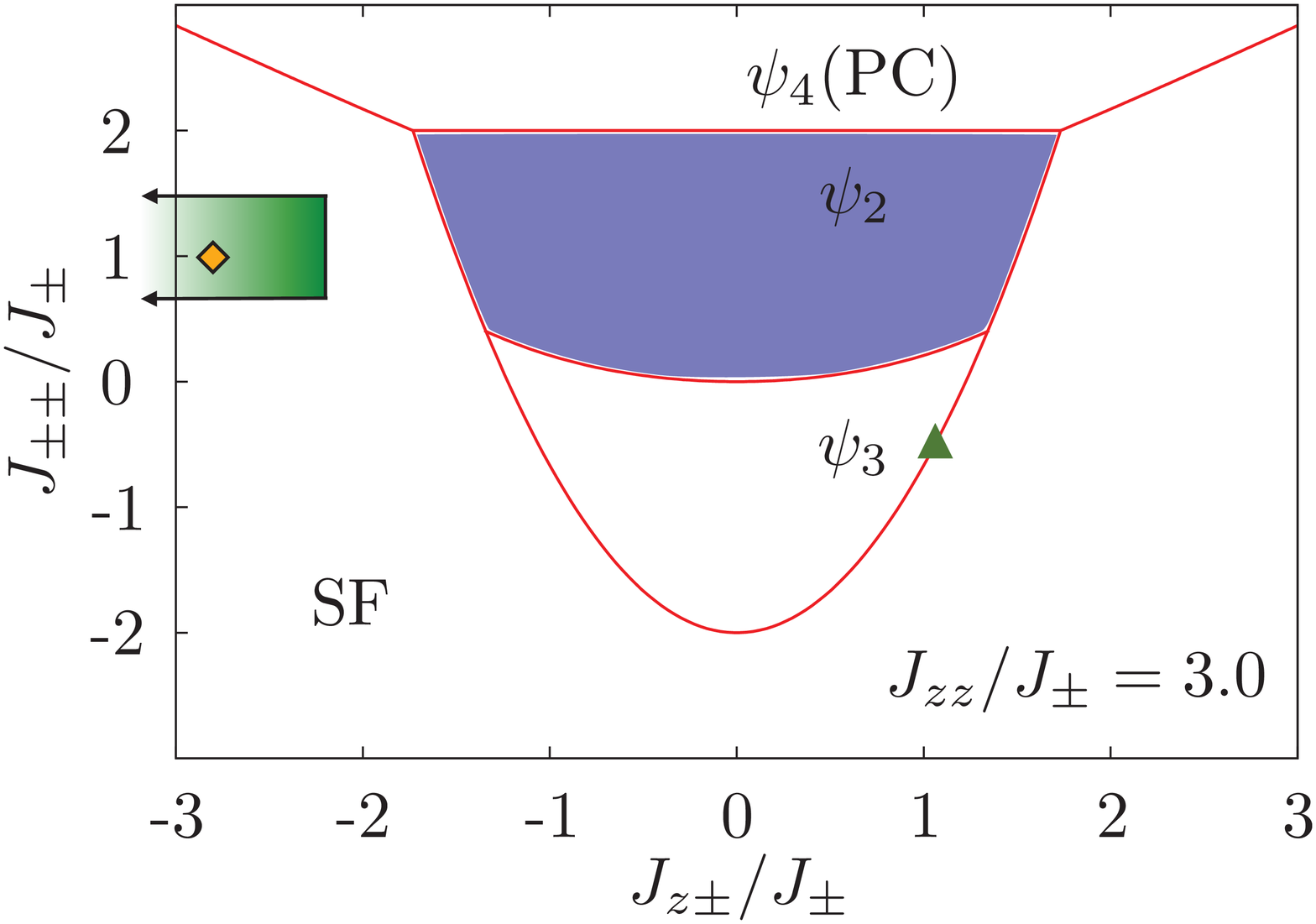}
\end{minipage}
\hfill
\begin{minipage}[b]{.33\textwidth}
\includegraphics[width=6cm]{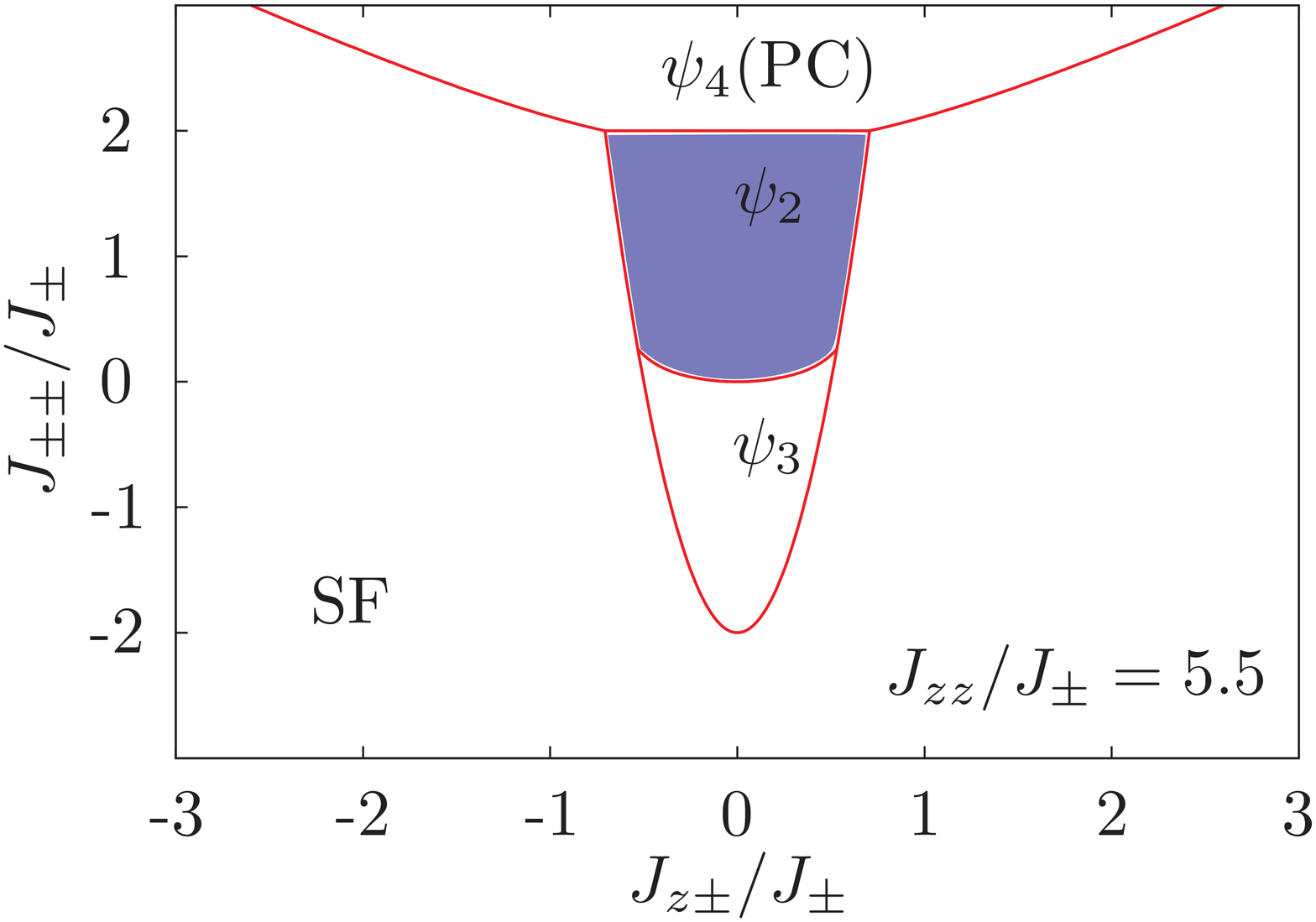}
\end{minipage}

\end{minipage}

\end{centering}

\caption{
Phase boundaries separating the $\psi_2$ and $\psi_3$ states for different values of  $j_{zz}\equiv J_{zz}/J_{\pm}$.
 The solid red lines are phase boundaries between the $\Gamma_5$ manifold and the Palmer-Chalker (PC, or
$\psi_4$ state of the $\Gamma_7$ irrep \cite{Poole.2007}, or the splayed ferromagnetic (SF) state, or,
within $\Gamma_5$, between the $\psi_2$ and $\psi_3$ states.
Within the $\Gamma_5$ manifold, $\psi_2$ is selected within the blue shaded regions.
The various symbols correspond to values of $J_e \equiv \{ J_{\pm}$, $J_{z\pm}$, $J_{\pm\pm}$ and $J_{zz} \}$ for pyrochlore  systems previously 
studied \cite{Ross.2011,Zhitomirsky.PhysRevLett.109.077204,Savary.PhysRevLett.109.167201,Chern.2010}.
 The $XY$ antiferromagnet model of Ref. \cite{Stasiak.2011} is at $j_{zz}=j_{z\pm}=0$ and $j_{\pm\pm}=2$, that is at the
$\psi_2$ (PC)  boundary where quantum fluctuations select $\psi_2$ \cite{Stasiak.2011,Zhitomirsky.PhysRevLett.109.077204}.
The red circle in $\psi_2$ ($J_{zz}/J_{\pm}=-0.5$ panel) is for Er$_2$Ti$_2$O$_7$ \cite{Savary.PhysRevLett.109.167201}.
Despite uncertainties in its $J_e$ couplings, we find that Er$_2$Ti$_2$O$_7$
 remains deeply in the $\psi_2$ region and does not cross in either $\psi_3$ or $\psi_4$ (PC).
The yellow squares in $\psi_2$ ($J_{zz}/J_{\pm}=0$ panel) are for the model of Ref.~[\onlinecite{Zhitomirsky.PhysRevLett.109.077204}] 
with anisotropic coupling ${j}_a\equiv {{\cal J}_a}/{\cal J}$  in their notation.
With $J_{\pm\pm}/J_{\pm}=(4-{j}_a)/(2+{j}_a)$, their model approaches the $\psi_2$ (PC) boundary as ${j}_a \rightarrow 0^+ $.
The green triangle ($J_{zz}/J_{\pm}=-0.5$, $J_{zz}/J_{\pm}=0$, $J_{zz}/J_{\pm}=0.5$, 
$J_{zz}/J_{\pm}=2.0$, $J_{zz}/J_{\pm}=3.0$ panels)
on the $\psi_3$/SF boundary is for the Heisenberg  pyrochlore antiferromagnet with 
indirect Dzyaloshinskii-Moriya interactions \cite{McClarty_ETO}  of varying strength studied 
in Ref.~[\onlinecite{Chern.2010}] which, coincidentally, resides on the $\Gamma_5$/SF boundary at the classical level (Appendix \ref{AA}).
The orange diamond corresponds to Yb$_2$Ti$_2$O$_7$ \cite{Ross.2011}. Because of the uncertainty in 
$J_{\pm}$, $J_{z\pm}$, $J_{\pm\pm}$ and $J_{zz}$, Yb$_2$Ti$_2$O$_7$  ``inhabits'' the $J_{zz}/J_{\pm}=2.0$ and $J_{zz}/J_{\pm}=3.0$ panels.
The green box delineates the ranges $J_{\pm\pm}/J_{\pm} \in [0.6,1.5]$ and
$J_{z\pm}/J_{\pm} \in [-3.8,-2.2]$, which extends well to the left of the vertical $J_{\pm\pm}/J_{\pm}$ axis, hence the left pointing arrows.
This figure illustrates that the Hamiltonian of  Yb$_2$Ti$_2$O$_7$ sits in the same SF phase recently reported for 
Yb$_2$Sn$_2$O$_7$ \cite{Yaouanc.PhysRevLett.110.127207}, far from the $\Gamma_5$ manifold
($\psi_2$ and $\psi_3$) as well as  the $\psi_4$ (PC) state.
}
\label{fig:diagram}
\end{figure*}

\section{Spin-wave calculations}
\label{sec:sw}

Apprised with the above  symmetry-constrained 
understanding of the phase diagram, we now present the results of our spin-wave calculations. 
We  employ the standard Holstein-Primakoff boson 
to represent fluctuations about long-range ordered  $\psi_2$ and $\psi_3$ states. 
Since the spins point along the local $x$ and $y$ direction in  $\psi_2$ or $\psi_3$, respectively (see Fig. \ref{psi2and3}),
the spin components in local coordinates are written as:
\begin{subequations}\label{hp}
\begin{eqnarray}
S_{i}^{(x,y)}&  \!  = \! &S-a_{i}^\dagger a_{i},  \\
S_{i}^{(y,z)}&  \! =  \! &\frac{\sqrt{2S}}{2}(a_{i}^\dagger+a_{i}),\\
S_{i}^{(z,x)}&\!= \!& \frac{\sqrt{2S}}{2i}(a_i-a_i^\dagger) .
%S_{i}^{(y,z)}&=&\frac{\sqrt{2S}}{2}(a_{i}^\dagger+a_{i}), 
%S_{i}^{(z,x)}&=&\frac{\sqrt{2S}}{2i}(a_i-a_i^\dagger) .
\end{eqnarray}
\end{subequations}
The superscripts $(\mu,\nu)$ of $S_i^{(\mu,\nu)}$ correspond to the boson representation for  spins in the ($\psi_2,\psi_3$) state, respectively.
 We substitute \eqref{hp} into \eqref{hami} and keep terms up to  quadratic order in $a_i$ and $a_i^\dagger$.
The bosons are written in term of Bloch modes and the spin-wave spectrum is determined in the usual way 
by a Bogoliubov transformation \cite{Maestro.2004}. We define the total zero-point energy for $\psi_{2,3}$ states, $E_{0}(\psi_{2,3})$ as: 
\begin{equation}\label{E0}
E_{0}(\psi_{i})\equiv \frac{1}{N}\sum_{\bm k}\sum_{\alpha=0}^{3}\frac{\hbar \omega_{i\alpha}({\bm k})}{2} ,
\end{equation}
where 
$\frac{1}{2}{\hbar \omega_{i\alpha}({\bm k})}$
 is the zero-point energy of the spin-wave mode of momentum ${\bm k}$  for $\psi_{i}$ state ($i=2,3$). 
$\alpha$ labels the spin-wave branches and the summation over ${\bm k}$ is restricted to the first Brillouin zone. 
$N$ is the number of primitive tetrahedra units.  Some $\omega_{i\alpha}({\bm k}=0)$ become imaginary 
if the interaction parameters puts the system outside the phase boundaries
 between the $\Gamma_5$ manifold and the  PC or SF long-range ordered states, 
which both have a ${\bm k}=0$ ordering wavevector.

We define $\delta E$  as $\delta E \equiv E_{0}(\psi_2) - E_{0}(\psi_3)$.
Positive or negative $\delta E$ signals that $\psi_3$ or $\psi_2$ is selected by quantum fluctuations, respectively. To check the symmetry arguments in Sec. \ref{mands}, 
we first calculate $\delta E(0,j_{\pm\pm})$ and $\delta E(j_{z\pm},0)$  for $j_{zz}=0$.
As shown in Fig. \ref{linescan},  $\delta E(0,j_{\pm\pm})\sim j_{\pm\pm}^3$ and $\delta E(J_{z\pm},0)\sim j_{z\pm}^6$, 
consistent with Eq. \eqref{cubic}. 

We then scan the  $j_{z\pm}$ and $j_{\pm\pm}$ parameter space for several values of $j_{zz}$.
The results are shown in Fig. \ref{fig:diagram}, which constitutes our main result.
There is only one phase boundary dividing regions of $\psi_2$ and $\psi_3$ 
states for large negative and positive $j_{zz}$. For intermediate values of $j_{zz}$, $0\lesssim j_{zz}\lesssim 3$ 
we observe {\it three phase boundaries} separating alternating regions of $\psi_2$ and $\psi_3$  .
 As $j_{zz} \rightarrow 3$ from below,  the (lower) narrow $\psi_2$ sliver region, 
sandwiched between two $\psi_3$ regions,
 expands and merges with the large $\psi_2$ region for $j_{zz}\sim 3$.
Also, the same narrow $\psi_2$ sliver disappears rapidly for $j_{zz} \lesssim 0$.
As expected, all phase boundaries touch at $j_{z\pm}=j_{\pm\pm}=0$ for arbitrary $j_{zz}$. 
No other phases but $\psi_2$ and $\psi_3$ were found stabilized by quantum fluctuations within the $\Gamma_5$ manifold.

It is appropriate at this point to comment on the role of higher order anisotropic terms in $\delta E(\phi)$. 
While we expect higher order corrections to $\delta E$ beyond $\cos(6\phi)$ of the form $\sum_{n>0} g_{2n+1}
\cos\{6(2n+1)\phi)\}$ (terms such as
$\cos(12n\phi)$ do not distinguish $\psi_2$ and $\psi_3$),
these do not lead to qualitative new behavior for the $J_{\pm\pm}/J_{\pm}$ vs
$J_{z\pm}/J_{\pm}$ phase diagram on the basis of our spin-wave calculations.
In particular, {\it no more} than three $\psi_2/\psi_3$ phase boundaries are observed in our explicit quantum
spin wave calculations, in agreement with the heuristic arguments leading to Eq.(4).
One does note in the insets of Fig. ~\ref{linescan}
deviations of $\delta E$, for either large $J_{z\pm}/J_\pm$ or $J_{\pm\pm}/J_\pm$, away
from the strict power laws ($\sim J_{z\pm}^6$ and $\sim J_{\pm\pm}^3$) behaviors expected on the basis of the lowest
$\cos(6\phi)$ harmomic (i.e. Eq. \eqref{cubic}).
Referring to Fig. 3, the roots of the Eq. \eqref{cubic},  $\delta E=0$,  would give for the
$\psi_2/\psi_3$ phase  boundaries
$j_{\pm\pm} = \omega_\mu (J_{zz}) j_{z\pm}^2$ (where $\mu=1,2,3$ and $\omega_\mu$ a real number).
This form would lead to a gap between the boundaries defining the narrow $\psi_2$
 sliver that would {\it monotonously} widen as $j_{z\pm}$ increases
if the cubic Eq. (4)  for $\delta E$ was exact.
However, as can be seen in Fig. 3   for $J_{zz}/J_\pm = -0.5$ and $J_{zz}/J_{\pm} = 0$, the width
of the sliver first grows as $J_{z\pm}/J_{\pm}$ increases, reaches a maximum width,
and then  {\it narrows} as the classical boundary $\Gamma_5/$SF boundary
given by Eq. (2) is approached from inside the $\Gamma_5$ region.
It therefore appears that the corrections to $\delta E$ beyond $\cos(6\phi)$ merely lead to a
slight renormalization of the ``internal''
(i.e. within the $\Gamma_5$ region) $\psi_2/\psi_3$ phase boundaries
defining the narrow $\psi_2$ sliver in the $J_{zz}/J_{\pm}=0$, $0.5$ and $2.0$ panels of Fig. 3.

\section{The role of long-range dipolar interaction in \eto}
\label{sec:dp}

Our knowledge of the general phase diagram in Fig \ref{fig:diagram} 
for anisotropic nearest-neighbor exchange 
$J_e = \{ J_\pm, J_{\pm\pm}, J_{z\pm}, J_{zz} \}$
motivates us  to return to an important material-relevant  question:
what is the role of the long-range $1/r^3$ magnetostatic dipole-dipole interaction in allowing for a $\psi_2$ ground state
induced by quantum ObD in Er$_2$Ti$_2$O$_7$\cite{Champion.PhysRevB.68.020401,Dalmas_ETO,Poole.2007}?  
The magnetic moment of Er$^{3+}$ in the local $XY$ plane is approximately 
3 $\mu_{\rm B}$ ~\cite{Champion.PhysRevB.68.020401}. As a result, the strength of 
the nearest-neighbor part of the 
dipolar interaction is of the order of 15\% to 50\% of the $J_e$ 
parameters determined in Ref.[\onlinecite{Savary.PhysRevLett.109.167201}]. 
It is therefore natural to ask whether the dipolar interaction beyond the
 nearest-neighbor affects the quantum ObD.
proposed to be at play in Er$_2$Ti$_2$O$_7$ \cite{Zhitomirsky.PhysRevLett.109.077204,Savary.PhysRevLett.109.167201}

To address this question, we adopt the values of 
$J_e = \{ J_{zz}, J_{\pm}, J_{\pm \pm}, J_{z \pm} \}$ 
with their uncertainty from Savary \emph{et al}.\cite{Savary.PhysRevLett.109.167201}.
 In order to avoid a double accounting of the nearest-neighbor contribution   {from} the dipoles implicitely contained 
in the experimentally determined
$J_e$, we need to subtract the nearest-neighbor contribution from the dipolar  
interaction from the $J_e$ determined in Ref. [\onlinecite{Savary.PhysRevLett.109.167201}].

%We the calculate the spin-wave spectrum including long-range dipolar interaction.
% The Fourier transformation of the long-range dipolar interaction is evaluated using Ewald summation. 
% The calculation is carried out within for a total of $3^6=729$ runs for the following range of couplings,  
% $J_{zz} = -2.5 \pm 1.8$, $J_{\pm} = 6.5 \pm 0.75$, $J_{\pm \pm} = 4.2 \pm 0.5$, $g_{xy}=5.97 \pm 0.08$,
% and $g_{z}=2.45 \pm 0.23$. Here $g_{xy}$ and $g_z$ are the perpendicular and longitudinal g-tensors with respect to the local $[111]$ direction. 
% Our calculation show that the ground state of Er$_2$Ti$_2$O$_7$ is $\psi_2$ with long-range dipolar interaction properly accounted. 

Working with bare couplings, the nearest-neighbor part of the dipolar Hamiltonian is:
\begin{equation}
\mathcal{H}^{\text{dip}}
=
D 
\displaystyle\sum\limits^{\text{}}_{<ab>}
\left[ \frac{ \mathbf{S}_{a} \cdot \mathbf{S}_{b} }{|\mathbf{R}_{ab}|^3} - 3 \frac{ \left( \mathbf{S}_{a} \cdot \mathbf{R}_{ab} \right) \left( \mathbf{S}_{b} \cdot \mathbf{R}_{ab} \right) }{|\mathbf{R}_{ab}|^5} \right]
\end{equation}
where
\begin{equation}
D = \frac{\mu_0 \mu^2_{\rm B}}{4 \pi}. 
\end{equation}
Here $\mu_0$ is the permeability of free space, $\mu_{\rm B}$ 
is the Bohr magneton, and $a_0 = 10.07$ \AA\ \cite{Knop.1965, Blote} is the conventional cubic unit cell lattice constant of \eto. 
The nearest-neighbor distance $|\mathbf{R}_{ab}| = a_0\sqrt{2}/4$.

Translating the dipolar interaction into a 
$\{ J^{\text{dip}}_{zz}, J^{\text{dip}}_{\pm}, J^{\text{dip}}_{\pm \pm}, J^{\text{dip}}_{z \pm} \}$
notation, we obtain the following couplings for the dipolar contribution to the nearest-neighbor interactions:
\begin{equation}
\begin{array}{ccc}
\left( \begin{array}{c}
J^{\text{dip}}_{zz} \\
J^{\text{dip}}_{\pm} \\
J^{\text{dip}}_{\pm \pm} \\
J^{\text{dip}}_{z \pm} \end{array} \right)
=
\frac{D}{12}
\left( \begin{array}{c}
20 {g_{z}}^2 \\
- {g_{xy}}^2 \\
7 {g_{xy}}^2 \\
- 2 \sqrt{2} {g_{xy}} {g_{z}} \end{array} \right)
\end{array}
\end{equation}
where $g_{xy}$ and $g_z$ are the perpendicular and longitudinal 
$g$-tensors with respect to the local $[111]$ direction, respectively.
 We subtract these dipolar contributions  from the experimental
$\{ J_{zz}, J_{\pm}, J_{\pm \pm}, J_{z \pm} \}$ couplings 
in Savary \emph{et al.}\cite{Savary.PhysRevLett.109.167201}. 

The nearest-neighbor dipole   couplings
we find for the best-fit couplings and 
$g$-tensors reported in Ref.~[\onlinecite{Savary.PhysRevLett.109.167201}]  are ($g$-tensored values in $10^{-2}$ meV)
\begin{equation}
\begin{array}{ccc}
\left( \begin{array}{c}
J^{\text{dip}}_{zz} \\
J^{\text{dip}}_{\pm} \\
J^{\text{dip}}_{\pm \pm} \\
J^{\text{dip}}_{z \pm} \end{array} \right)
=
\left( \begin{array}{c}
1.18997 \\
-0.353283 \\
2.47298 \\
-0.410071 \end{array} \right) . 
\end{array}
\label{Jdip}
\end{equation}
These values   {differ slightly} from the values 
$\left( 0.8, -0.46, 3.2, -0.38\right)$ reported in the Supplementary material of Ref.~[\onlinecite{Savary.PhysRevLett.109.167201}]
  {as} the authors of that paper
did not use their   {own} best fitted  $g$-tensor values   {for their calculation of the $J^{\text{dip}}_{uv}$} values.

We next proceed to calculate the spin-wave spectrum including long-range dipolar interaction following the method of Ref.~[\onlinecite{Maestro.2004}]
using either $\psi_2$ or $\psi_3$ as reference (degenerate) classical ground state.
The Fourier transformation of the long-range dipolar interaction is evaluated using Ewald summation \cite{Enjalran.2004,Maestro.2004}
The calculation is carried out  for a total of $3^6=729$ sets 
of couplings and $g$-tensor values within the following ranges reported in 
Ref.~[\onlinecite{Savary.PhysRevLett.109.167201}]:
$J_{zz} = -2.5 \pm 1.8$, $J_{\pm} = 6.5 \pm 0.75$, 
$J_{\pm \pm} = 4.2 \pm 0.5$, $g_{xy}=5.97 \pm 0.08$, and $g_{z}=2.45 \pm 0.23$.  
For {\it all} $729$ combinations of these six quantities, 
our  {calculations} show that the ground state of Er$_2$Ti$_2$O$_7$ 
is  $\psi_2$ when the long-range dipolar interactions are
properly accounted   {for}. 

While the $J^{\text{dip}}_{uv}$ couplings in Eq. (\ref{Jdip}) are   {not} small compared to the
experimental $J^{\text{e}}_{uv}$ determined in Ref.~[\onlinecite{Savary.PhysRevLett.109.167201}], the conclusion that quantum order-by-disorder 
into $\psi_2$ is operational for Er$_2$Ti$_2$O$_7$ is   {unchanged} because the material ``resides'' deeply in the $\psi_2$ state 
of  the phase diagram of Fig. 3 ($J_{zz}/J_\pm = -0.5$ panel).
That being said, we would expect that consideration of the 
long-range dipolar interaction would renormalize the spin-wave gap 
$-$ a necessary signature of the broken discrete symmetry $\psi_2$ state~\cite{Stasiak.2011} 
$-$ computed in Ref.~[\onlinecite{Savary.PhysRevLett.109.167201}].

 The conclusion of Ref.~[\onlinecite{Savary.PhysRevLett.109.167201}] that quantum ObD is responsible for the $\psi_2$ ground state for
the nearest-neighbor $J_e$ exchange parameters determined by inelastic neutron scattering is thus upheld. 
This is the main result of our work as per the ground state of \eto. 

\section{Discussion}
\label{sec:discuss}

In this work,we studied the quantum order-by-disorder for a
pyrochlore XY magnet with $J_e = \{ J_{\pm},J_{z\pm},J_{\pm\pm},J_{zz} \}$
exchange couplings between effective spin-1/2 degrees of freedom. 
We determined the region of Je interaction parameters where the
classically degenerate $\Gamma_5$ manifold is the ground state by
minimizing the classical energy. Within the 
$\Gamma_5$ manifold, the boundaries between the $\psi_2$ and $\psi_3$ states are obtained by
comparing the contribution of zero-point energy from spin
waves to the total energy of the system.
We recover the results of several previous works  
\cite{Stasiak.2011,Zhitomirsky.PhysRevLett.109.077204,Savary.PhysRevLett.109.167201,Chern.2010} 
as special cases of our study (see caption of Fig.~\ref{fig:diagram}).

We exposed that there can be one or three phase boundaries depending on the value of $J_{zz}$.
 This observation was anticipated since the number of $\psi_2/\psi_3$ phase boundaries 
is accurately controlled by the number of real solutions of the cubic Eq.~\eqref{cubic}.
While the exact location of the phase boundaries are expected to shift if interactions between spin-waves
are included or the temperature is finite, the topology of the phase diagram is, however, 
 governed by the  symmetry arguments presented in this paper. 
Guided by these arguments, it would be interesting to explore  how 
the phase boundaries between $\psi_2$ and $\psi_3$  states evolve with temperature.
Such a phenomenon was studied in Ref.~[\onlinecite{Chern.2010}] where $\psi_2$ order below and near the 
critical temperature changes to $\psi_3$ order at lower temperature.
We expect such a scenario to occur in materials close to one of the $\psi_2/\psi_3$ zero temperature 
phase boundaries.
%% We expect such scenarios can be realized if relevant materials are close to 
%% the phase boundary of $\psi_2$ and $\psi_3$ states at zero temperature.
%% Our phase diagram should provide guidance to search for multiple phase transitions 
%%in  $XY$ pyrochlore materials, such as Er$_2$Ge$_2$O$_7$ which belongs to 
%%the recently synthesized family of rare-earth (R) R$_2$Ge$_2$O$_7$ pyrochlore compounds \cite{R2Ge2O7_PRL}.
 In addition, we speculate that Er$_2$Sn$_2$O$_7$ may be close to one of the $\psi_2/\psi_3$ phase boundaries,
where the selection of either state can become weak, 
or close to the $\psi_2/$PC  boundary near $J_{\pm\pm}/J_\pm=2$.
 This could help explain why 
%rationalize the experimental observation that  
Er$_2$Sn$_2$O$_7$ fails to order down to 50 mK \cite{Lago_ETO,Sartre_ETO,Guitteny.2013}. 

Other possibilities for exploring the physics of $XY$ pyrochlore magnets may include
Er$_2$Ge$_2$O$_7$ and Yb$_2$Ge$_2$O$_7$ \cite{Hallas.2013}.
%http://meetings.aps.org/Meeting/MAR13/Session/T15.13
It may also be possible that some of the aforementioned 
phases may be realized in materials other than the $R_2M_2$O$_7$ pyrochlore oxides \cite{Gardner_RMP}.
For example, the Cd$R_2$Se$_2$ and Cd$R_2$S$_2$ chalcogenide spinel compounds, 
in which the $R$ trivalent rare-earth ions sit on a regular pyrochlore lattice of corner-sharing tetrahedra,
provide a new opportunity for the study of frustrated magnetism on the pyrochlore lattice \cite{Lau.PhysRevB.72.054411}.
By rescaling the crystal field parameters determined for the spinel-based spin 
ice compound CdEr$_2$Se$_4$ \cite{Lago.PhysRevLett.104.247203}, we predict (Appendix \ref{Sect-4}) that the Dy$^{3+}$ ions in
the sister compound CdDy$_2$Se$_4$ would be $XY$-like and could possibly 
realize some of the interesting physics described above.

Given  the rich phase diagram of Fig. \ref{fig:diagram}, especially that of the $\Gamma_5$ manifold,
it would be interesting to investigate whether transitions between $\psi_2$ and $\psi_3$ states can be driven by various control 
parameters. This includes external magnetic field along different symmetry directions \cite{Ruff_ETO}, hydrostatic
 \cite{Mirebeau_TTO,Mirebeau_TSO}
or chemical pressure \cite{R2Ge2O7_PRL}
 as well as random disorder, being either ``intrinsic'' and caused by magnetic rare-earth ions 
substituting for the transition metal ion (e.g Ti$^{4+}$, dubbed as ``stuffing'') 
\cite{Ross_stuff} or ``extrinsic'' 
and driven by diluting the Er$^{3+}$ sites by non-magnetic Y$^{3+}$ ions, for example \cite{Ke_PRL,Lin.2013}.
With the deeper understanding of quantum ObD in $XY$ pyrochlore magnets reached in the present work, we
hope that our work will stimulate further systematic experimental and theoretical studies of $XY$ pyrochlore oxides and spinel materials.

\section*{ACKNOWLEDGEMENT}

M.G. acknowledges Ludovic Jaubert for a stimulating
discussion regarding the problem of degeneracy and order-bydisorder
in $XY$ pyrochlore magnets.We thank Alexandre Day,
Behnam Javanparast, Jaan Oitmaa and Rajiv Singh for useful
discussions and related collaborations and acknowledge Boris
Malkin for useful correspondence. We thank Taoran Lin for
help with the figures. This work was supported by the NSERC
of Canada, the Canada Research Chair program (M.G., Tier
1), and the Perimeter Institute (PI) for Theoretical Physics.
Research at PI is supported by the Government of Canada
through Industry Canada and by the Province of Ontario
through theMinistry of Economic Development \& Innovation.

\appendix
\section{TRANSLATION BETWEEN DIFFERENT CONVENTIONS}
\label{AA}

The  Hamiltonians of Ref.~[\onlinecite{Zhitomirsky.PhysRevLett.109.077204}] and [\onlinecite{Chern.2010}] are, respectively:
\begin{subequations}
\begin{eqnarray}
\mathcal{H}_{\text{Z}} &=& \sum_{\langle ij\rangle}
\left[ 
\left( \mathcal{J}  + \frac{\mathcal{J}_{\text{a}}}{3} \right) \mathbf{S}^{\perp}_{i} \cdot \mathbf{S}^{\perp}_{j} 
-\frac{\mathcal{J}_{\text{a}}}{3} \left( \mathbf{S}^{\perp}_{i} \cdot \mathbf{S}^{\perp}_{j}\right.\right.\nonumber\\ &&\left.\left. -
3 \left( \mathbf{S}^{\perp}_{i} \cdot \mathbf{\hat{r}}_{ij} \right) \left( \mathbf{S}^{\perp}_{j} \cdot \mathbf{\hat{r}}_{ij} \right) \right)
\right],\\\
\mathcal{H}_{\text{C}}& = &\sum_{\langle ij\rangle}
\left[ 
J \left( \mathbf{S}_{i} \cdot \mathbf{S}_{j} \right)
+ D \left( \mathbf{\hat{\Omega}}^{\text{DM}}_{ij} \cdot \mathbf{S}_{i} \times \mathbf{S}_{j} \right)
\right]. 
\end{eqnarray}
\end{subequations}
Here the superscript $\perp$ denotes the component perpendicular to the local [111] axes (i.e. the XY model). $\hat{r}_{ij}$ is the unit vector pointing from site $i$ to site $j$. $D$ is the strength of Dzyaloshinskii-Moriya (DM) interaction. $\mathbf{\hat{\Omega}}^{\text{DM}}_{ij}$ is the DM vector  \cite{Elhajal.PhysRevB.71.094420} on bond $\langle ij\rangle$ such that $D>0$ corresponds to \emph{indirect} DM interaction. 
% Note that the coupling $\mathcal{J}_{\text{a}}$ is a combination of $\mathcal{J}_{\text{iso}}^{\perp}$ and $\mathcal{J}_{\text{pd}}^{\perp}$.
We first translate $\mathcal{J}$ and $\mathcal{J}_{\text{a}}$ 
in ${\cal H}_{\rm Z}$ of Ref.~[\onlinecite{Zhitomirsky.PhysRevLett.109.077204}] 
above:
\begin{equation}
J_{\pm}=\frac{\mathcal{J}}{6}+\frac{\mathcal{J}_{\text{a}}}{12},\qquad J_{\pm\pm}=\frac{\mathcal{J}}{3}-\frac{\mathcal{J}_{\text{a}}}{12},\qquad J_{zz}=J_{z\pm}=0. 
\end{equation}
As a result, $j_{\pm\pm}=(4\mathcal{J}-\mathcal{J}_{\text{a}})/(2\mathcal{J}+\mathcal{J}_{\text{a}})$ and $j_{zz}=j_{z\pm}=0$. 

\vspace{2mm}

We note here that had Ref.~[\onlinecite{Zhitomirsky.PhysRevLett.109.077204}] considered $J_{\text{a}} >4$, we
predict from the results in Fig. 3 that they would have observed order-by-disorder  in $\psi_3$.

\vspace{5mm}

We now translate $J$ and $D$ in the Hamiltonian ${\cal H}_{\rm C}$ of Ref.~[\onlinecite{Chern.2010}] above:
\begin{equation}
\begin{aligned}
&J_{\pm}=\frac{J}{6}+\frac{\sqrt{2}D}{6},\qquad J_{\pm\pm}=\frac{J}{3}-\frac{\sqrt{2}D}{6},\\
&J_{zz}=-\frac{J}{3}+\frac{2\sqrt{2}D}{3},\qquad J_{z\pm}=\frac{\sqrt{2}J}{3}+\frac{D}{6}. 
\end{aligned}
\end{equation}
Consequentially, we have:
\begin{subequations}
\begin{eqnarray}
j_{\pm\pm}&=&\frac{2J-\sqrt{2}D}{J+\sqrt{2}D}, \\
j_{zz}&=&\frac{-2J+4\sqrt{2}D}{J+\sqrt{2}D},\\
j_{z\pm}&=&\frac{2\sqrt{2}J+D}{J+\sqrt{2}D}.
\end{eqnarray} 
\end{subequations}
For $0<D<\infty$, $-2<j_{zz}<4$. 

Below we define $\gamma_{ij}$ and $\zeta_{ij}$ in the convention of Ref.~[\onlinecite{Ross.2011}]. The pyrochlore lattice has four sublattices $0$, $1$, $2$ and $3$. Both $\gamma_{ij}$ and $\zeta_{ij}$ are labeled by the sublattices site $i$ and $j$ belong to. 
\begin{subequations}
\begin{eqnarray}
&&\gamma_{01}=-\zeta_{01}^*=\gamma_{23}=-\zeta_{23}^*=1,\\ &&\gamma_{02}=-\zeta_{02}^*=\gamma_{13}=-\zeta_{13}^*=\exp\left(\frac{2\pi i }{3}\right),\\ &&\gamma_{03}=-\zeta_{03}^*=\gamma_{12}=-\zeta_{12}^*=\exp\left(-\frac{2\pi i}{3}\right).
\end{eqnarray}
\end{subequations}

\section{EXPECTED EFFECTIVE {\it XY} SPIN FOR 
D{\lowercase{y}}$^{3+}$ in C{\lowercase{d}}D{\lowercase{y}}$_2$S{\lowercase{e}}$_4$ SPINEL}
\label{Sect-4}

Recent work \cite{Lago.PhysRevLett.104.247203} has shown that in CdEr$_2$Se$_2$, the Er$^{3+}$ 
can be described by a classical Ising spin (with $g_{zz}$ = 16.05, and $g_{xy}=0$), 
with the system displaying a residual Pauling entropy and is thus a spin ice compound \cite{Gardner_RMP,LMM}.
The effective pseudospin-1/2 anisotropy, or components of the 
$g$-tensors, of a ground-state doublet are determined from the crystal field Hamiltonian,
${\cal H}_{\rm cf}$, acting on the rare-earth ion.
Ref.~[\onlinecite{Lago.PhysRevLett.104.247203}] provides an estimate of the parameters that determine ${\cal H}_{\rm cf}$ for CdEr$_2$Se$_2$.
This, in return, allows one to roughly estimate the crystal field parameters of other rare-earth ions in a same isotructural familly.
Here, we are particularly interested in CdDy$_2$Se$_4$ studied in Ref.~[\onlinecite{Lau.PhysRevB.72.054411}].

The crystal field Hamiltonian, ${\cal H}_{\rm cf}$ acting on a rare-earth can  be written as
\begin{equation}
{\cal H}_{\rm cf}= \sum_i \sum_{l,m} \tilde B_l^m O_l^m( {\bm J}_i)      \; ,
\label{Hcf}
\end{equation}
where the $\tilde B_l^m$ are the crystal field parameters and the $O_l^m(J_i^z,J_i^\pm)$  are the Stevens equivalent
operators functions of the components $J_i^z$ and $J_i^\pm$ of the angular momentum \cite{Stevens,Hutchings}.

In the so-called Stevens formalism, $\tilde B_2^0 \approx \alpha_J  \langle {\rm r}^2\rangle A_2^0$,
where $A_2^0$ is a point charge lattice sum representing the crystal-field, 
 $\langle {\rm r}^2\rangle$ the expectation value of ${\rm r}^2$ for the 4f electrons, 
and where we neglected the so-called charge shielding factor \cite{Gardner_RMP}. 
As discussed in Ref.~[\onlinecite{Gardner_RMP}], the 
$\alpha_J$ Stevens factor~\cite{Hutchings,Stevens} 
changes sign in a systematic pattern throughout the lanthanide series.
It is  positive for Sm, Er, Tm and Yb and negative for all others.
So, the sign of $\tilde B_2^0$ depends on the product $\alpha_J A_2^0$.
As Ref.~[\onlinecite{Lago.PhysRevLett.104.247203}] 
finds $A_2^0<0$\,\cite{Diff_CEF_notation}  for Er$^{3+}$ in CdEr$_2$Se$_4$, 
a key factor for making Er$^{3+}$ Ising-like in this compound, we thus 
expect $\tilde B_2^0$ to be positive for Dy$^{3+}$ in CdDy$_2$Se$_4$ \cite{Lau.PhysRevB.72.054411},
thus suggesting that Dy$^{3+}$ would be XY-like in CdDy$_2$Se$_4$.

To investigate this suggestion further, we take the $A_l^m$ coefficients ($B_l^m$ in 
the notation of Ref.~[\onlinecite{Lago.PhysRevLett.104.247203,Diff_CEF_notation}]) and, 
using the Stevens factors for Er$^{3+}$ and Dy$^{3+}$ \cite{Hutchings},
 the $\langle {\rm r}^n\rangle$ radial expectation values for
Er$^{3+}$ and Dy$^{3+}$ taken from Ref.~[\onlinecite{Freeman}], 
we can estimate the $\tilde B_l^m$ in Eq.~(\ref{Hcf})
for CdDy$_2$Se$_4$.
We find
$\tilde B_2^0 =  -260.95$,
$\tilde B_4^0 = -957.65$,
$\tilde B_4^3 = -1139.5$,
$\tilde B_6^0 = 167.93$,
$\tilde B_6^3 = -256.97$ and 
$\tilde  B_6^6 = 142.96$,
all in K units.
We then proceed to diagonalize ${\cal H}_{\rm cf}$, finding a ground state doublet with $g$-tensor components 
$g_{zz}= 3.36$ and $g_{xy} = 8.19$, with an energy gap of $\Delta = 18.1$ K from the lowest energy excited doublet.

With $g_{xy}/g_{zz} \sim 2.4$, we are thus led to anticipate that the CdDy$_2$Se$_4$ \cite{Lau.PhysRevB.72.054411}
 might constitute an interesting 
realization of magnetic system described by a 
pseudospin-1/2 Hamiltonian such as Eq. (1) 
in the main text, and whose thermodynamic properties may be rationalized on the basis of the discussion presented in the paper.
In particular, being an $XY$ system, it may display a phenomenon of  quantum order-by-disorder 
if its exchange parameters $J_e$ position it in the $\Gamma_5$ phase diagram of Fig. (3) in the main text.
Alternatively, depending on its $J_e$ couplings, it may find itself in either a $\psi_4$ or SF phase or even allow for a more
exotic possibility, and display a $U(1)$ quantum spin liquid or Coulomb ferromagnet as discussed in Ref.~[\onlinecite{Savary.PhysRevLett.108.037202}].

\bibliography{eto_PRB}

\begin{thebibliography}{44}
\bibitem[{but()}]{but_excited_CEF}
\bibinfo{note}{See, however, Ref.~[\onlinecite{McClarty_ETO}] for a discussion
  of a weak energetic mechanism operating at the mean-field level and proceding
  via the excited crystal field states.}
\end{thebibliography}
\end{document}